\documentclass[a4paper,11pt]{article}
\usepackage{pos}

\usepackage{slashed,stmaryrd,bm}
\usepackage{multirow}
\newcommand{\Dlr}{\mbox{$\raisebox{2mm}{\boldmath ${}^\leftrightarrow$}\hspace{-4mm}D^{}_\mu$}}
\newcommand{\Dilr}{\mbox{$\raisebox{2mm}{\boldmath ${}^\leftrightarrow$}\hspace{-4mm}D^I_\mu$}}

\newcommand{\Yn}{Y^{}_\nu}
\newcommand{\DYn}{Y^\dagger_\nu}
\newcommand{\Yl}{Y^{}_l}

\newcommand{\Lelli}[1][\beta]{\ell^{}_{#1 \rm L}}
\newcommand{\BLelli}[1][\alpha]{\overline{\ell^{}_{#1 \rm L}}}
\newcommand{\REi}[1][\beta]{E^{}_{#1 \rm R}}
\newcommand{\BREi}[1][\alpha]{\overline{E^{}_{#1 \rm R}}}
\newcommand{\LQi}[1][\beta]{Q^{}_{#1 \rm L}}
\newcommand{\BLQi}[1][\alpha]{\overline{Q^{}_{#1 \rm L}}}
\newcommand{\RUi}[1][\beta]{U^{}_{#1 \rm R}}
\newcommand{\BRUi}[1][\alpha]{\overline{U^{}_{#1 \rm R}}}
\newcommand{\RDi}[1][\beta]{D^{}_{#1 \rm R}}
\newcommand{\BRDi}[1][\alpha]{\overline{D^{}_{#1 \rm R}}}

\newcommand{\rmI}{{\rm i}}

\newcommand{\Op}{\mathcal{O}}

\newcommand{\str}[1]{{\rm STr} \left( #1 \right)}

\title{One-loop Matching for the Seesaw Model}

\author*[a,b]{Di Zhang}
\author[a,b]{Shun Zhou}

\affiliation[a]{
	Institute of High Energy Physics, Chinese Academy of Sciences, Beijing 100049, China}

\affiliation[b]{School of Physical Sciences, University of Chinese Academy of Sciences, Beijing 100049, China}

\emailAdd{zhangdi@ihep.ac.cn}
\emailAdd{zhoush@ihep.ac.cn}

\abstract{In this talk, we present the results of the complete one-loop matching of the seesaw model onto its low-energy effective theory by integrating out three right-handed neutrinos at the one-loop level. We find that there are 31 independent dimension-6 (dim-6) operators (barring flavor structures and Hermitian conjugates) in the Warsaw basis, and the standard-model couplings and the Wilson coefficient of the Weinberg operator acquire threshold corrections at the one-loop level. The complete Lagrangian up to dim-6 at the one-loop level is derived, which is indispensable for consistent explorations of the impact of heavy right-handed neutrinos at low energies.}

\FullConference{%
  41st International Conference on High Energy physics - ICHEP2022\\
  6-13 July, 2022\\
  Bologna, Italy
}

\begin{document}
\maketitle

\section{Introduction}

The canonical seesaw model with three right-handed neutrinos \cite{seesaw} is one of the most natural and the simplest extensions of the Standard Model (SM) to account for tiny but nonzero neutrino masses. As a bonus, the seesaw model also provides an elegant explanation for the matter-antimatter asymmetry of the Universe via the leptogenesis mechanism \cite{Fukugita:1986hr}. Since the mass scale of right-handed neutrinos is typically much higher than the electroweak scale, i.e.,  $M \gg \Lambda^{}_{\rm EW} \sim \mathcal{O} \left( 10^2 \right)$ GeV, one can integrate out heavy right-handed neutrinos and work in the corresponding effective field theory (EFT), i.e., the so-called seesaw EFT (SEFT) to explore low-energy consequences of heavy right-handed neutrinos. In this way, one can easily perform large logarithm resummations by means of the renormalization group equations (RGEs) in the EFT. The tree-level structure of SEFT derived by integrating out right-handed neutrinos at the tree level has been obtained in Ref. \cite{Broncano:2002rw}, and contains the unique Weinberg operator \cite{Weinberg:1979sa} and two dimension-six (dim-6) operators in the  Warsaw basis. The former one generates tiny neutrino masses, whereas the latter two modify the couplings of neutrinos with weak gauge bosons after gauge symmetry is spontaneously broken. However, if one wants to take into account one-loop effects in the seesaw model, the tree-level SEFT is not enough. In this case, the one-loop SEFT achieved by integrating out heavy right-handed neutrinos at the one-loop level is indispensable for consistently exploring one-loop consequences of right-handed neutrinos, e.g., radiative decays of charged leptons in the SEFT \cite{Zhang:2021tsq}. In this talk, we summarize the results of one-loop matching of the seesaw model, i.e., integrate out heavy right-handed neutrinos at the one-loop level, and achieve the complete one-loop structure of the SEFT \cite{Zhang:2021jdf}. Then, we show self-consistent calculations of radiative decays of charged leptons.

\section{Matching between ultraviolet (UV) theory and EFT}

The idea to match a given UV theory to the corresponding low-energy EFT is to equate the one-light-particle-irreducible (1LPI) effective action (i.e., $\Gamma^{}_{\rm L, UV}$) in the UV theory with the one-particle-irreducible (1PI) effective action (i.e., $\Gamma^{}_{\rm EFT}$) in the EFT at the matching scale $\mu$, i.e.,
\begin{eqnarray}
	\Gamma^{}_{\rm L, UV} \left[ \phi^{}_{\rm B}\right] = \Gamma^{}_{\rm EFT} \left[ \phi^{}_{\rm B} \right] \;,
\end{eqnarray}
in which $\phi^{}_{\rm B}$ denotes the light background field and both effective actions can be calculated by means of the background field method \cite{Abbott:1981ke}. By making use of the functional approach, one can easily derive the tree-level Lagrangian of the EFT, namely
\begin{eqnarray}\label{eq:tree-level}
	\mathcal{L}^{\rm tree}_{\rm EFT} \left[ \phi^{}_{\rm B} \right] = \mathcal{L}^{}_{\rm UV} \left[ \hat{\Phi}_{\rm c} \left[ \phi^{}_{\rm B} \right] , \phi^{}_{\rm B} \right] \;,
\end{eqnarray}
where $\hat{\Phi}^{}_{\rm c} \left[ \phi^{}_{\rm B} \right]$ is the localized solution of the classical equation of motion for the heavy field $\Phi^{}_{\rm B}$, 
\begin{eqnarray}
\left. 	\frac{\delta \mathcal{L}_{\rm UV} \left[ \Phi, \phi \right] }{\delta \Phi} \right|^{}_{\Phi = \Phi^{}_{\rm c} \left[ \phi^{}_{\rm B} \right], \phi = \phi^{}_{\rm B}} = 0 \;.
\end{eqnarray}
The one-loop Lagrangian of the EFT can be calculated via \cite{Cohen:2020fcu}
\begin{eqnarray}\label{eq:one-loop}
	\int {\rm d}^d x \mathcal{L}^{\rm 1-loop}_{\rm EFT} \left[ \phi^{}_{\rm B} \right] = \left. \frac{\rmI}{2} {\rm STr} \ln \left( - \bm{K } \right)  \right|^{}_{\rm hard} - \left. \frac{\rmI}{2} \sum^\infty_{n=1} \frac{1}{n} {\rm STr} \left[ \left( \bm{ K^{-1} X} \right)^n \right] \right|^{}_{\rm hard} \;,
\end{eqnarray}
where $\bm{K}$ and $\bm{X}$ are the inverse-propagator and interaction matrices, respectively, and can be extracted from the UV Lagrangian via
\begin{eqnarray}
	\left. \delta^2 \mathcal{L}^{}_{\rm UV} \right|^{}_{\Phi = \hat{\Phi}^{}_{\rm c} \left[ \phi^{}_{\rm B} \right]} = \left. 2\mathcal{L}^{}_{\rm UV} \left[ \varphi + \delta \varphi \right] \right|^{}_{\Phi = \hat{\Phi}^{}_{\rm c} \left[ \phi^{}_{\rm B} \right]} \supset \delta \overline{\varphi}^{}_i \left( \bm{K}^{}_i \delta^{}_{ij} - \bm{X}^{}_{ij}  \right) \delta \varphi^{}_j
\end{eqnarray}
with $\varphi$ being field multiplet and containing fields and their conjugates. Then, one may calculate supertraces in Eq. \eqref{eq:one-loop} with the help of the Mathematica package {\sf SuperTracer} \cite{Fuentes-Martin:2020udw} or {\sf STrEAM} \cite{Cohen:2020qvb} based on the covariant derivative expansion (CDE) method \cite{Gaillard:1985uh}.

\section{Complete one-loop structure of the SEFT}

\begin{table}[t!]
	\centering
	\renewcommand\arraystretch{1.5}
	\resizebox{\textwidth}{!}{
		\begin{tabular}{c|c|c|c|c|c}
			\hline\hline
			\multicolumn{2}{c}{$X^2H^2$} & \multicolumn{2}{|c|}{$\psi^2DH^2$} & \multicolumn{2}{c}{Four-quark}
			\\
			\hline
			$\Op^{}_{HB}$ & $B^{}_{\mu\nu} B^{\mu\nu} H^\dagger H$ & $\Op^{(1)\alpha\beta}_{HQ}$ &  $\left( \BLQi \gamma^\mu \LQi \right) \left( H^\dagger \rmI~ \Dlr H \right)$ & $\Op^{(1)\alpha\beta\gamma\lambda}_{qu}$ & $\left( \BLQi \gamma^\mu \LQi \right) \left( \BRUi[\gamma] \gamma^{}_\mu \RUi[\lambda] \right)$
			\\
			$\Op^{}_{HW}$ & $W^I_{\mu\nu} W^{I\mu\nu} H^\dagger H$ & $\Op^{(3)\alpha\beta}_{Hq} $ & $ \left( \BLQi \gamma^\mu \tau^I \LQi \right) \left(H^\dagger \rmI~ \Dilr H \right)$ & $\Op^{(8)\alpha\beta\gamma\lambda}_{qu}$ & $\left(\BLQi \gamma^\mu T^A \LQi \right) \left( \BRUi[\gamma] \gamma^{}_\mu T^A \RUi[\lambda] \right)$
			\\
			$\Op^{}_{HWB}$ & $W^I_{\mu\nu} B^{\mu\nu} \left( H^\dagger \tau^I H \right)$ & $\Op^{\alpha\beta}_{Hu}$ & $\left( \BRUi \gamma^\mu \RUi \right) \left(H^\dagger \rmI~ \Dlr H \right)$ & $ \Op^{(1)\alpha\beta\gamma\lambda}_{qd}$ & $ \left( \BLQi \gamma^\mu \LQi \right) \left( \BRDi[\gamma] \gamma^{}_\mu \RDi[\lambda] \right)$
			\\
			\cline{1-2}
			\multicolumn{2}{c|}{$H^4D^2$} & $\Op^{\alpha\beta}_{Hd}$ & $\left( \BRDi \gamma^\mu \RDi \right) \left( H^\dagger \rmI ~\Dlr H \right)$ & $\Op^{(8)\alpha\beta\gamma\lambda}_{qd}$ & $\left( \BLQi \gamma^\mu T^A \LQi \right) \left( \BRDi[\gamma] \gamma^{}_\mu T^A \RDi[\lambda] \right)$
			\\
			\cline{1-2}
			$\Op^{}_{H\Box}$ & $\left(H^\dagger H\right) \Box \left( H^\dagger H \right)$ & $\Op^{(1)\alpha\beta}_{H\ell}$ & $\left( \BLelli \gamma^\mu \Lelli \right) \left( H^\dagger \rmI ~\Dlr H \right)$ & $\Op^{(1)\alpha\beta\gamma\lambda}_{quqd}$ & $ \left( \overline{Q^a_{\alpha{\rm L}}} \RUi \right) \epsilon^{ab} \left( \overline{Q^b_{\gamma{\rm L}}} \RDi[\lambda] \right) $
			\\
			\cline{5-6}
			$\Op^{}_{HD}$ & $\left(H^\dagger D^{}_\mu H \right)^\ast \left(H^\dagger D^\mu H \right)$ & $\Op^{(3)\alpha\beta}_{H\ell}$ & $ \left( \BLelli \gamma^\mu \tau^I \Lelli \right) \left( H^\dagger \rmI ~\Dilr H \right) $ & \multicolumn{2}{c}{Four-lepton}
			\\
			\cline{1-2}\cline{5-6}
			\multicolumn{2}{c|}{$H^6$} & $\Op^{\alpha\beta}_{He}$ & $\left(\BREi \gamma^\mu \REi \right) \left( H^\dagger \rmI~ \Dlr H \right)$ & $\Op^{\alpha\beta\gamma\lambda}_{\ell\ell}$ & $ \left( \BLelli \gamma^\mu \Lelli \right) \left( \BLelli[\gamma] \gamma^{}_\mu \Lelli[\lambda] \right) $
			\\
			\cline{1-4}
			$\Op^{}_H$ & $\left(H^\dagger H \right)^3$ & \multicolumn{2}{c|}{$\psi^2 H^3$} & $\Op^{\alpha\beta\gamma\lambda}_{\ell e}$ & $\left(\BLelli \gamma^\mu \Lelli \right) \left( \BREi[\gamma] \gamma^{}_\mu \REi[\lambda] \right) $
			\\
			\cline{1-4}
			\multicolumn{2}{c|}{$\psi^2XH$} & $\Op^{\alpha\beta}_{uH}$ & $\left( \BLQi \widetilde{H} \RUi \right) \left( H^\dagger H \right) $ & &
			\\
			\cline{1-2}
			$\Op^{\alpha\beta}_{eB}$ & $\left( \BLelli \sigma^{\mu\nu} \REi \right) H B^{}_{\mu\nu}$ & $\Op^{\alpha\beta}_{dH}$ & $ \left( \BLQi H \RDi \right) \left( H^\dagger H \right) $ & &
			\\
			$\Op^{\alpha\beta}_{eW}$ & $ \left( \BLelli \sigma^{\mu\nu} \REi \right) \tau^I H W^I_{\mu\nu} $ & $\Op^{\alpha\beta}_{eH}$ & $ \left( \BLelli H \REi \right) \left( H^\dagger H \right)$ & &
			\\
			\hline
			\multicolumn{6}{c}{Semi-leptonic}
			\\
			\hline
			$\Op^{(1)\alpha\beta\gamma\lambda}_{\ell q}$ & $ \left( \BLelli \gamma^\mu \Lelli \right) \left( \BLQi[\gamma] \gamma^{}_\mu \LQi[\lambda] \right)$ & $\Op^{\alpha\beta\gamma\lambda}_{\ell u}$ & $\left( \BLelli \gamma^\mu \Lelli \right) \left( \BRUi[\gamma] \gamma^{}_\mu \RUi[\lambda] \right)$ & $\Op^{\alpha\beta\gamma\lambda}_{\ell e d q}$ & $ \left( \BLelli \REi \right) \left( \BRDi[\gamma] \LQi[\lambda] \right) $
			\\
			$\Op^{(3)\alpha\beta\gamma\lambda}_{\ell q}$ & $ \left( \BLelli \gamma^\mu \tau^I \Lelli \right) \left( \BLQi[\gamma] \gamma^{}_\mu \tau^I \LQi[\lambda] \right) $ & $\Op^{\alpha\beta\gamma\lambda}_{\ell d}$ & $\left( \BLelli \gamma^\mu \Lelli \right) \left( \RDi[\gamma] \gamma^{}_\mu \RDi[\lambda] \right)$ & $\Op^{(1)\alpha\beta\gamma\lambda}_{\ell e q u}$ & $\left( \overline{\ell^a_{\alpha \rm L}} \REi \right) \epsilon^{ab} \left( \overline{Q^b_{\gamma \rm L}} \RUi[\lambda] \right)$
			\\
			\hline\hline
	\end{tabular}}
	\vspace{-0.15cm}
	\caption{Dim-6 operators in the seesaw model at the one-loop level in the Warsaw basis, where the Hermitian conjugates of the operators in classes $\psi^2 X H$ and $\psi^2 H^3$ and four-fermion operators are not listed explicitly.}
	\label{tb:Wbasis}
	\vspace{-0.25cm}
\end{table}

With the help of Eqs. \eqref{eq:tree-level} and \eqref{eq:one-loop}, and the Lagrangian of the seesaw model, i.e.,
\begin{eqnarray}\label{eq:Lagrangian}
	\mathcal{L}^{}_{\rm UV} = \mathcal{L}^{}_{\rm SM} + \overline{N^{}_{\rm R}} \rmI \slashed{\partial} N^{}_{\rm R} - \left( \frac{1}{2} \overline{N^c_{\rm R}} MN^{}_{\rm R} + \overline{\ell^{}_{\rm L}} Y^{}_\nu \widetilde{H} N^{}_{\rm R} + {\rm h.c.} \right) 
\end{eqnarray}
with $\mathcal{L}^{}_{\rm SM}$ being the SM Lagrangian, one can obtain the complete Lagrangian of the SEFT up to dim-6 at the one-loop level, namely,
\begin{eqnarray}
	\mathcal{L}^{}_{\rm SEFT} &=& \mathcal{L}^{}_{\rm SM} \left( m^2 \to m^2_{\rm eff}, \lambda \to \lambda^{}_{\rm eff}, Y^{}_l \to Y^{\rm eff}_l, Y^{}_{\rm u} \to Y^{\rm eff}_{\rm u}, Y^{}_{\rm d} \to Y^{\rm eff}_{\rm d}  \right)
	\nonumber
	\\
	&& + \left[ \frac{1}{2} \left( C^{(5)}_{\rm eff} \right)_{\alpha\beta} \Op^{(5)}_{\alpha\beta} + {\rm h.c.} \right] + \frac{1}{4} \left( C^{(6)}_{\rm tree} \right)_{\alpha\beta} \left[ \Op^{(1)\alpha\beta}_{H\ell} - \Op^{(3)\alpha\beta}_{H\ell} \right] + \sum_i C^{}_i \Op^{}_{i} \;,
\end{eqnarray}
where $\Op^{}_i$ denote the dim-6 operators listed in Table~\ref{tb:Wbasis} including Hermitian conjugations of the non-Hermitian operators, while $C^{}_i$ refer to the one-loop contributions to corresponding Wilson coefficients. Explicit expressions of all the effective couplings and Wilson coefficients can be found in Ref. \cite{Zhang:2021jdf}. Here, we only show those for $\Op^{}_{eB}$ and $\Op^{}_{eW}$, i.e., 
\begin{eqnarray}
	C^{\alpha\beta}_{eB} = \frac{g^{}_1}{24 \left( 4\pi \right)^2} \left( \Yn M^{-2} \DYn \Yl \right)^{}_{\alpha\beta} \;,\quad C^{\alpha\beta}_{eW} = \frac{5g^{}_2}{24\left( 4 \pi \right)^2} \left( \Yn M^{-2} \DYn \Yl \right)^{}_{\alpha\beta} \;.
\end{eqnarray}
These two dim-6 operators directly contribute to radiative decays of charged leptons after spontaneous gauge symmetry breaking as follows \cite{Zhang:2021tsq}
\begin{eqnarray}\label{eq:rad}
	\rmI \mathcal{M} \left( l^{-}_\beta \to l^{-}_\alpha + \gamma \right) = \frac{\rmI e g^2_2}{6\left( 4\pi \right)^2 M^2_W} \left( RR^\dagger \right)^{}_{\alpha\beta} \left[ \epsilon^\ast_\mu \overline{u} \left( p^{}_2 \right) \rmI \sigma^{\mu\nu} q^{}_\nu \left( m^{}_\alpha P^{}_{\rm L} + m^{}_\beta P^{}_{\rm R} \right) u \left( p^{}_1 \right) \right]
\end{eqnarray}
where $R \equiv vY^{}_\nu M^{-1}/\sqrt{2}$ with $v\approx 246~{\rm GeV}$ and $m^{}_\alpha$ denotes the mass of $l^-_\alpha$. Together with one-loop contributions from the tree-level SEFT, Eq. \eqref{eq:rad} reproduces the results in the full theory \cite{Cheng:1980tp}.

\section{Summary}

We have carried out the complete one-loop matching of the seesaw model onto the corresponding SEFT, and obtained 31 independent dim-6 operators (barring flavor structures and Hermitian conjugates) in the Warsaw basis up to the one-loop level. Moreover, the SM couplings acquire threshold corrections in the SEFT, which are the very matching conditions for two-loop RGEs. The obtained one-loop structure of the SEFT is indispensable to consistent calculations of one-loop consequences of heavy right-handed neutrinos in the seesaw model.
\\[0.12cm]
{\it This work was supported by the National Natural Science Foundation of China under grants No. 12075254 and No. 11835013.}


\begin{thebibliography}{99}
\bibitem{seesaw}
P.~Minkowski,
Phys. Lett. B \textbf{67} (1977), 421-428;
T.~Yanagida,
Conf. Proc. C \textbf{7902131} (1979), 95-99
KEK-79-18-95;
M.~Gell-Mann, P.~Ramond and R.~Slansky,
Conf. Proc. C \textbf{790927} (1979), 315-321
[arXiv:1306.4669 [hep-th]];
S.~L.~Glashow,
NATO Sci. Ser. B \textbf{61} (1980), 687;
R.~N.~Mohapatra and G.~Senjanovic,
Phys. Rev. Lett. \textbf{44} (1980), 912.
\\[-0.8cm]
\bibitem{Fukugita:1986hr}
M.~Fukugita and T.~Yanagida,
Phys. Lett. B \textbf{174} (1986), 45-47.
\\[-0.8cm]
\bibitem{Broncano:2002rw}
A.~Broncano, M.~B.~Gavela and E.~E.~Jenkins,
Phys. Lett. B \textbf{552} (2003), 177-184
[erratum: Phys. Lett. B \textbf{636} (2006), 332];
Nucl. Phys. B \textbf{672} (2003), 163-198.
\\[-0.8cm]
\bibitem{Weinberg:1979sa}
S.~Weinberg,
Phys. Rev. Lett. \textbf{43} (1979), 1566-1570.
\\[-0.8cm]
\bibitem{Zhang:2021tsq}
D.~Zhang and S.~Zhou,
Phys. Lett. B \textbf{819} (2021), 136463.
\\[-0.8cm]
\bibitem{Zhang:2021jdf}
D.~Zhang and S.~Zhou,
JHEP \textbf{09} (2021), 163.
\\[-0.8cm]
\bibitem{Abbott:1981ke}
L.~F.~Abbott,
Acta Phys. Polon. B \textbf{13} (1982), 33
CERN-TH-3113.
\\[-0.8cm]
\bibitem{Cohen:2020fcu}
T.~Cohen, X.~Lu and Z.~Zhang,
JHEP \textbf{02} (2021), 228.
\\[-0.8cm]
\bibitem{Gaillard:1985uh}
M.~K.~Gaillard,
Nucl. Phys. B \textbf{268} (1986), 669-692;
L.~H.~Chan,
Phys. Rev. Lett. \textbf{57} (1986), 1199;
O.~Cheyette,
Nucl. Phys. B \textbf{297} (1988), 183-204.
\\[-0.8cm]
\bibitem{Fuentes-Martin:2020udw}
J.~Fuentes-Martin, M.~K\"onig, J.~Pag\`es, A.~E.~Thomsen and F.~Wilsch,
JHEP \textbf{04} (2021), 281.
\\[-0.8cm]
\bibitem{Cohen:2020qvb}
T.~Cohen, X.~Lu and Z.~Zhang,
SciPost Phys. \textbf{10} (2021) no.5, 098.
\\[-0.8cm]
\bibitem{Cheng:1980tp}
T.~P.~Cheng and L.~F.~Li,
Phys. Rev. Lett. \textbf{45} (1980), 1908;
A.~Ilakovac and A.~Pilaftsis,
Nucl. Phys. B \textbf{437} (1995), 491;
R.~Alonso, M.~Dhen, M.~B.~Gavela and T.~Hambye,
JHEP \textbf{01} (2013), 118;
Z.~z.~Xing and D.~Zhang,
Eur. Phys. J. C \textbf{80} (2020) no.12, 1134.

\end{thebibliography}
\end{document}